\documentclass[a4paper,11pt,UKenglish]{article}

\usepackage{amssymb,amsmath,natbib,amsthm}
\usepackage{graphicx,ifpdf,color} 
\usepackage[bottom=1in, left=1in, right=1in]{geometry}
\usepackage[figuresright]{rotating}
\usepackage{subcaption}

\usepackage{bm}

\newcommand{\Sidak}{\v Sid\'ak }

\ifpdf 
   \DeclareGraphicsExtensions{.pdf,.eps,.png} 
\fi

\begin{document}

\title{Permutation in genetic association studies with covariates: controlling the familywise error rate with score tests in generalized linear models}

\author{K. K. Halle$^{1,2}$ and M. Langaas$^{1}$\\ \\
\footnotesize\parbox{.88\linewidth}{$^1$Department of Mathematical Sciences, NTNU, Norwegian University of Science
and Technology, NO-7491 Trondheim, Norway.
$^2$Liaison Committee between the Central Norway Regional Health Authority (RHA) and the Norwegian University of Science and Technology (NTNU), Trondheim, Norway.}
}
\date{\today}
\maketitle

\begin{abstract}

In genome-wide association (GWA) studies the goal is to detect associations between genetic markers 
and a given phenotype. The number of genetic markers can be large and effective methods for control of the overall error rate is 
a central topic when analyzing GWA data. The Bonferroni method is known to be conservative when the tests are dependent. Permutation
methods give exact control of the overall error rate when the assumption of exchangeability is satisfied, but are computationally intensive for large
datasets. For regression models the exchangeability assumption is in general not satisfied and
there is no standard solution on how to do permutation testing, except some approximate methods. 
In this paper we will discuss permutation methods for control of the familywise error rate in genetic association studies and present an approximate solution. These methods will be compared using simulated data. 
\end{abstract}


\section{Introduction}\label{sec:introduction}

In genome-wide association (GWA) studies, genetic markers are tested one at a time for association with a given phenotype.
The number of markers is large ($\sim 10^5-10^6$), and we need efficient methods for multiple testing correction to control the overall error rate.
We consider single-step multiple testing procedures, which control the overall error rate by estimating one local significance level, $\alpha_{\text{loc}}$, 
to be used as the cut-off value for detecting significance for each individual test. 

The Bonferroni method gives strong control of the familywise error rate (FWER) for all types of dependence structures, but is
known to be conservative when the tests are dependent. The \Sidak method assumes that the test statistics are independent and gives strong control of the FWER. 
Resampling methods can be used when the parametric distribution of a test statistic 
is unknown or requires complicated formulas (for example high dimensional integrals) to compute. 
The two main types of resampling methods are permutation and bootstrap methods. The maxT permutation method 
of \cite{WestfallYoung1993} controls the FWER when the assumption of exchangeability is satisfied.  This assumption is in general not satisfied for generalized linear models. Background theory about exchangeability is given in \cite{Commenges2003}. 

\cite{Halle2016} presented an alternative to the Bonferroni method for multiple
testing correction in generalized linear models. This method is based on using the asymptotic multivariate normal distribution of the score test statistics and approximates
high dimensional integrals by several integrals of lower dimension. In this paper, we consider methods for approximating high dimensional integral by resampling methods. 

Permutation methods for a single hypothesis test modelled by linear models have been studied by \cite{FreedmanLane1983}, \cite{TerBraak1992} 
and \cite{Renaud2010} among others. The main approaches for permutation methods in normal linear models are to consider either permuting the raw data
or to permute the residuals of the model. An comparison of some of the resampling methods is found in \cite{Anderson1999} and \cite{Anderson2001}. The 
methods of \cite{FreedmanLane1983} and \\ \cite{Renaud2010} are based on permuting the residuals under a reduced model, while the method of \cite{TerBraak1992} 
is based on permuting the residuals under the full model. Permuting the residuals under the reduced or the full model will have asymptotically exact significance levels
\citep{Anderson1999}. The method of \cite{FreedmanLane1983} is based on a linear model and the method of \cite{Renaud2010} is used in the ANOVA setting. In this paper these methods will be adapted to and compared in the multiple testing setting. 

For single and multiple hypothesis testing with generalized linear models, there is no standard solution on how to do permutation testing. In this paper, we will give a review of permutation methods for control of the FWER for normal linear models. We will discuss the assumption of exchangeability and present an approximate solution for permutation testing in generalized linear models. 

This paper is organized as follows. In Section \ref{sec:background} we will present some concepts and theoretical background
for resampling methods and in Section \ref{sec:lm} we will present some existing methods for permutation and $p$-value estimation
in normal linear models. In Section \ref{sec:glm} we will present permutation methods for generalized linear models and multiple testing correction.
The methods are compared in Section \ref{sec:results}. The paper will conclude with discussion and conclusion in Section \ref{sec:discussion}. 

\section{Statistical background}\label{sec:background}

In this section we present the score test for generalized linear models, 
the concepts of subset pivotality and exchangeability and some basic theory about methods to correct for multiple testing. 

\subsection{Notation}\label{sec:notation}

We assume that we have data - one phenotype, $m$ genetic markers and $d$ environmental covariates from $n$ independent individuals. 
Let $\bm{Y}$ be a $n$-dimensional vector with the response variable. Let $X_{\text{e}}$ be a $n \times d$ matrix of environmental covariates (intercept in the first column), and $X_{\text{g}}$ 
a $n \times m$ matrix of genetic markers, then $X = (X_{\text{e}} X_{\text{g}})$ is a $n \times (d+m)$ covariate matrix. The genetic data are assumed to be from common variant biallelic genetic markers
with alleles $a$ and $A$, where $A$ is assumed to be the minor allele, based on the estimated minor allele frequency. We use additive coding $0,1,2$ for the three possible genotypes
$aa, Aa$ and $AA$, respectively. 

\subsection{The score test}\label{sec:scoretest}

The $n$ individuals are assumed to be independent and the phenotype for the $i$'th individual is denoted by $Y_i, i=1,\ldots,n$, where
\begin{align*}
\text{E}(Y_i) = \mu_i \text{ and } \text{Var}(Y_i)=\sigma_i^2. 
\end{align*}

We assume that the relationship between the $n$-dimensional vector of phenotypes, $\bm{Y}$, and the covariate matrix, $X$, can be modeled using a generalized linear model (GLM) \citep[Chapter~2]{McCullaghNelder1989}, with a $n$-dimensional
vector of linear predictors
\begin{equation*}
\bm{\eta} = X_{\text{e}}\bm{\beta}_{\text{e}} + X_{\text{g}}\bm{\beta}_{\text{g}} = X\bm{\beta},
\end{equation*}
where $\bm{\beta}=(\bm{\beta}_{\text{e}}^T \bm{\beta}_{\text{g}}^T)^T$ is a $(d+m)$-dimensional parameter vector. 
The score vector to be used for testing the null hypothesis $H_0: \bm{\beta}_{\text{g}}=\bm{0}$ is given by
\begin{align*}
\bm{U}_{\text{g}|\text{e}} = \frac{1}{\phi}X_{\text{g}}^T(\bm{Y}-\hat{\bm{\mu}}_{\text{e}}) 
\end{align*}
where $\phi$ is the dispersion parameter, $\hat{\bm{\mu}}_{\text{e}}$ is the fitted values from the null model with only the environmental covariates, $X_{\text{e}}$, present. The residual vector, $\hat{\bm{\epsilon}}$, is $\hat{\bm{\epsilon}}=\bm{Y}-\hat{\bm{\mu}}_{\text{e}}$. For $Y_i$ normally distributed, the dispersion parameter is $\phi=\sigma^2$ and for 
$Y_i$ Bernoulli distributed, $\phi=1$. The vector $\bm{U}_{\text{g}|\text{e}}$ is asymptotically normally distributed with mean $\bm{0}$ and covariance matrix $V_{\text{g}|\text{e}}=\frac{1}{\phi^2}(X_{\text{g}}^T\Lambda X_{\text{g}}-X_{\text{g}}^T\Lambda X_{\text{e}}(X_{\text{e}}^T \Lambda X_{\text{e}})^{-1}X_{\text{e}}^T \Lambda X_{\text{g}})$ (see \cite{Smyth2003}), where $\Lambda$ is a diagonal matrix with $\text{Var}(Y_i)$ on the diagonal.

We are not interested in testing the complete null hypothesis $H_0: \bm{\beta}_{\text{g}}=\bm{0}$, instead we are interested in testing the null hypothesis
$H_{0j}: \beta_{\text{g}j}=0$ for each genetic marker $j, j=1,\ldots, m$. Let $\bm{U}_{\text{g}|\text{e} \text{ } j}$ be the $j$'th component of the score vector $\bm{U}_{\text{g}|\text{e} }$ and 
$V_{\text{g}|\text{e} \text{ } jj}$ be element $jj$ of the matrix $V_{\text{g}|\text{e}}$. 
We consider the standardized components of the score vector, $\bm{T}=(T_1, \ldots, T_m)$, where
\begin{align}
T_j = \frac{\bm{U}_{\text{g}|\text{e} \text{ } j}}{\sqrt{V_{\text{g}|\text{e} \text{ } jj}}}, j = 1,\ldots, m. 
\label{Tk}
\end{align}
Note that the dispersion parameter $\phi$ is canceled in the test statistics, but the elements of $\Lambda$ need to be estimated.
Each component $T_j, j=1,\ldots, m$ is asymptotically standard normally distributed and the vector $\bm{T}$ is asymptotically 
multivariate normally distributed, $\bm{T} \sim N_m(\bm{0}, R)$, where the elements of the covariance matrix $R$ is 
$\text{Cov}(T_i,T_j)=\frac{V_{\text{g}|\text{e} \text{ } ij}}{\sqrt{V_{\text{g}|\text{e} \text{ } ii}V_{\text{g}|\text{e} \text{ } jj}}}$. 
We define $X_{\text{g}j}^T$ to be the $j$'th row of $X_{\text{g}}$ and write the score test statistic for the $j$'th hypothesis as
\begin{align}
T_j 		&= \frac{X_{\text{g}j}^T\hat{\bm{\epsilon}}}{\sqrt{X_{\text{g}j}^T\Lambda^{1/2}(I_n-H_{\Lambda})\Lambda^{1/2}X_{\text{g}j}}}
\label{eq:Tscore}
\end{align}
where  $I_n$ is the $n \times n$ identity matrix and 
\begin{align}
H_{\Lambda}=\Lambda^{1/2}X_{\text{e}}(X_{\text{e}}^T \Lambda X_{\text{e}})^{-1}X_{\text{e}}^T \Lambda^{1/2}
\label{hatmatrix}
\end{align} is the matrix which projects onto the column space of $\Lambda^{1/2}X_{\text{e}}$. 
The vector of score test statistics is $\bm{T}=(T_1,\ldots,T_m)$, where $T_j$ is given by Equation \eqref{eq:Tscore}.

\subsection{Multiple hypothesis testing}\label{sec:multtest}

We consider a multiple testing problem where each of $m$ genetic markers are tested for association with the phenotype. 
The unobserved number of erroneously rejected null hypotheses are denoted by $V$. The FWER is defined as the probability 
of at least one false positive result
\begin{align*}
\text{FWER}=P(V>0),
\end{align*}
and we consider methods which control the FWER at level $\alpha$. 
For each genetic marker $j, j = 1,\ldots, m$, we perform a score test, testing the null hypothesis, $H_{0j}: \beta_{\text{g}j}=0$, of no association between the 
genetic marker and the phenotype. The $p$-values, $p_j, j = 1, \ldots, m$ are the lowest nominal levels to reject $H_{0j}$.  

We consider single-step multiple testing methods, which use a so-called local significance 
level, $\alpha_{\text{loc}}$, as the cut-off value for detecting significance. For these methods, all hypotheses with a
$p$-value below $\alpha_{\text{loc}}$ will be rejected. If the local significance level, $\alpha_{\text{loc}}$, yields $\text{FWER} \leq \alpha$, we define the multiple testing method as valid. 

The Bonferroni method estimates the local significance level, $\alpha_{\text{loc}}$, by
$\alpha_{\text{loc}}=\frac{\alpha}{m}$ and gives strong control of the FWER for all types of dependence structures between the test statistics, but is known to be conservative when 
the tests are dependent. Strong control of the FWER means control of the FWER under any combination of true and false null hypotheses \citep{Goeman2014}. 

The \Sidak method assumes the tests are independent and estimates the local significance level by $\alpha_{\text{loc}}=1-(1-\alpha)^{1/m}$.
The \Sidak method also gives strong control of the FWER. 

Following the notation in \cite{Halle2016}, for each genetic marker $j, j =1,\ldots, m$ the event $O_j: |T_j| < c$, is the event where the null hypothesis for genetic marker $j$ is not rejected and the probability of the complementary event $\bar{O}_j$ is $P(\bar{O}_j)=2\Phi(-c)=\alpha_{\text{loc}}$.
The FWER can then be written as 
\begin{align}
\text{FWER}=1-P(O_1 \cap \cdots \cap O_m). 
\label{fwer}
\end{align}

When the vector of test statistics asymptotically follows a multivariate normal distribution as in Section \ref{sec:scoretest}, the joint probability in Equation \eqref{fwer} will be a $m$-dimensional integral
in a multivariate normal distribution. The method of \cite{Halle2016} approximate the high dimensional 
integral $P(O_1 \cap \cdots \cap O_m)$ by several integrals of low dimension. Another solution to estimate the FWER or calculate the local significance level, $\alpha_{\text{loc}}$, is to approximate the high dimensional integral by permutation methods, such as the maxT permutation method of \cite{WestfallYoung1993}. 
When the number of genetic markers is small, the high dimensional integral can also be solved using numerical integration methods, for example the method by \cite{genz1992,genz1993}, which is implemented for $m\leq 1000$ in the R package mvtnorm \citep{mvtnorm}.

\subsection{The maxT permutation procedure}\label{sec:maxT}
We consider the maxT permutation method described by \cite{WestfallYoung1993}. We write Equation \eqref{fwer} as
\begin{align}
\text{FWER}=P(\max_{j=1,\ldots,m} |T_j| \geq c).
\label{eq:maxT}
\end{align}

The maxT permutation method is based on estimating the empirical distribution of the maximal test statistic by
resampling the data under the complete null hypothesis, thus the exchangeability assumption needs to be satisfied, see Section \ref{sec:exchangeability}.
If the exchangeability assumption is satisfied \citep{Commenges2003}, we may use the empirical distribution of the maximal test statistic to estimate the cut-off value $c$ as in Equation \eqref{eq:maxT}.  Assume $B$ permutations of the data is performed and 
let $\bm{T}_b$ be the vector of score test statistics based on the $b$'th permutation of the data. 
Then $c$ is estimated to be the largest value where
\begin{align}
\frac{\#(\max |\bm{T}_b| \geq c)+1}{B+1} \leq \alpha.
\label{c}
\end{align}
If we assume $\bm{T}_b \sim N(0,1)$, the local significance level is found to be $\alpha_{\text{loc}}=2\Phi(-c)$.
For a given cut-off value $c$ we may use permutation methods to estimate the FWER by
\begin{align}
\hat{\alpha}=\frac{\#(\max |\bm{T}_b| \geq c)+1}{B+1}
\end{align}

\subsubsection{Subset pivotality}

We consider resampling methods for control of the FWER, where the set of true null hypotheses are unknown. 
The subset pivotality property was described for resampling methods by \citet[p.~42]{WestfallYoung1993}.
When the subset pivotality assumption is satisfied, we may resample the data under the complete null hypothesis and get strong control of the FWER, which means control of the FWER under any combination of true and false hypotheses \citep{Goeman2014}.

The subset pivotality property is satisfied if the joint distribution of the test statistics corresponding to the true null hypotheses does not depend on the distribution of the remaining test statistics \citep[p.~42]{WestfallYoung1993}. An intersection hypothesis is an hypothesis where two or more of the null hypotheses, $H_{0j}$, are tested simultaneously. We define $H_I$ to be the set of all possible intersection hypotheses, $H_I = \cap_{j \in I} H_{0j} $
where $I$ is all possible subsets of $\{1,\ldots,m\}$. When the subset pivotality condition is satisfied, the distribution of $\max_{j \in I} |T_j|$ and $\max_{j \in I} |T_j|$ 
are identical under the intersection hypothesis $H_I$ and under the complete null hypothesis for all intersection hypotheses. 
From Equation \eqref{eq:Tscore} we see that the score test statistic for a given genetic marker $j$ does not depend on the other 
genetic markers, and therefore, the subset pivotality condition is satisfied for the multiple testing problem using the GLM and score test statistics. When we have subset pivotality, we also have strong control of the FWER.


\subsubsection{Exchangeability}\label{sec:exchangeability}

The term exchangeability was introduced by de Finetti in the 1930s and is a key assumption of 
permutation methods. The vector $\bm{Y}$  has an exchangeable distribution
if and only if any permutation of the vector $\bm{Y}$ has the same distribution as $\bm{Y}$ \citep{Commenges2003}. 
A permutation matrix $P$ is a $n \times n$ matrix with elements 0 and 1, only. The matrix $P$ has exactly one entry of 1 in each row and each column and 0 elsewhere. 
For a permutation matrix $P$, we have $P^TP=I$. If $P$ is a $n \times n$ permutation matrix,  exchangeability is defined as \citep{Commenges2003}
\begin{align}
\bm{Y} \overset{d}{=} P\bm{Y} \quad \text{under } H_0
\label{ydist}
\end{align}
where $\overset{d}{=}$ means equality in distribution. Other forms of exchangeability also exist. If the distribution of $\bm{Y}$ and $P\bm{Y}$ are 
equal up to the second moment, then $\bm{Y}$ is second moment exchangeable. \\

Following the notation in  \cite{Commenges2003}, we write the vector of test statistics as a function, $f$, of the data, $\bm{Y}$, $\bm{T}=f(\bm{Y})$. If $\bm{Y}$ is not exchangeable, 
we find a transformation, $\tilde{\bm{Y}}=V(\bm{Y})$ of the data, such that $\tilde{\bm{Y}}$ is exactly or for example second moment exchangeable \citep{Commenges2003} and $\bm{T}=f(\bm{Y})=\tilde{f}(\tilde{\bm{Y}})$. We estimate the distribution of the maximal 
test statistic based on permutations of $\tilde{\bm{Y}}$.

\section{Permutation methods for regression models}\label{sec:resamplingmethods}
 
The maxT permutation method presented in Section \ref{sec:maxT} is based on the assumption of 
exchangeability, which in general is not satisfied for generalized linear models. For the 
normal linear model, a review of some approximate solutions for single hypothesis testing will be presented in Section \ref{sec:lm}. These methods are also set into our multiple testing problem. 
In Section \ref{sec:glm} we present resampling methods for generalized linear models.

\subsection{Permutation methods for the normal linear model}\label{sec:lm}

Permutation methods for testing a single hypothesis in the normal linear model have been discussed by \cite{FreedmanLane1983}, \cite{TerBraak1992} and \cite{Renaud2010}
among others. The main approaches for permutation testing for the normal linear model are to resample the raw data or to resample the residuals under either the
full model \citep{TerBraak1992}, a reduced model \citep{FreedmanLane1983} or a modified model \citep{Renaud2010}. A comparison of these methods 
are found in \cite{Anderson2001} among others. In this section, we will present the methods of \cite{FreedmanLane1983}, \cite{TerBraak1992} and \cite{Renaud2010}, 
and relate the methods to our score test statistic and the maxT permutation method with the aim to control the FWER. 

\subsubsection{Permute the residuals under the reduced model}\label{sec:ImH}
We consider the linear model
\begin{align}
\bm{Y}=X_{\text{e}}\bm{\beta}_{\text{e}}+X_{\text{g}}\bm{\beta}_{\text{g}}+\bm{\epsilon}
\label{lm}
\end{align}
where $\text{E}(\bm{\epsilon})=0$ and $\text{Cov}(\bm{\epsilon})=\sigma^2I_n$.

The method of \cite{FreedmanLane1983} is based on permuting the residuals of a reduced model, a model eliminating the nuisance parameters from the model in Equation \eqref{lm}.  
For the linear model this can be done by projecting the model in Equation \eqref{lm} onto the subspace orthogonal to the subspace spanned by the columns of $X_{\text{e}}$. This can
be done by multiplying both sides of the model with the projection matrix $(I_n-H_{\Lambda})$ where $I_n$ is the $n \times n$ identity matrix, and $H_{\Lambda}$ 
is the regression hat matrix as in Equation \eqref{hatmatrix}. For the normal linear model, the hat matrix equals $H_{\Lambda}=X_e(X_e^TX_e)^{-1}X_e^T$. This defines the residual vector, $\hat{\bm{\epsilon}}$, and gives the relationship
\begin{align}
\hat{\bm{\epsilon}}=(I_n-H_{\Lambda})\bm{Y} = (I_n-H_{\Lambda})X_{\text{g}}\bm{\beta}_{\text{g}}+(I_n-H_{\Lambda})\bm{\epsilon},
\label{model2}
\end{align}
and under $H_0: \bm{\beta}_{\text{g}}=\bm{0}$
\begin{align} 
\text{E}(\hat{\bm{\epsilon}})=\bm{0} \text{ and } \text{Cov}(\hat{\bm{\epsilon}})=\sigma^2I_n.
\label{emodel}
\end{align}
Using the results in Appendix \ref{sec:hatmatrix}, 
\begin{align*}
\text{Cov}(\hat{\bm{\epsilon}})=\text{Cov}((I_n-H_{\Lambda})\bm{Y})=(I_n-H_{\Lambda})\sigma^2, 
\end{align*}
and when $n \rightarrow \infty$ and the data contains no leverage points it can be shown that $(I_n-H_{\Lambda})\sigma^2 \rightarrow \sigma^2$ \citep[p.~207]{Brenton}. Thus, the residuals $\hat{\bm{\epsilon}}=(I_n-H_{\Lambda})\bm{Y}$ are asymptotically second moment exchangeable. 
The score test statistics for this model are given in Equation \eqref{eq:Tscore} and by permuting the
residuals, $\hat{\bm{\epsilon}}$, we can estimate the distribution of the maximal test statistic, and then 
estimate the local significance level, $\alpha_{\text{loc}}$. Let $P$ be a $n \times n$ permutation matrix.
The permuted score score test statistic for the genetic marker $j$ in the $b$'th permutation is
\begin{align}
\bm{T}_{\text{b}j} = \frac{X_{\text{g}j}^TP(I_n-H_{\Lambda})\bm{Y}}{\sqrt{X_{\text{g}j}^T\Lambda^{1/2}(I_n-H_{\Lambda})\Lambda^{1/2}X_{\text{g}j}}}.
\label{freedmanlane}
\end{align}
The vector of score test statistics for the $b$'th permutation is $\bm{T}_{\text{b}}=(\bm{T}_{\text{b}1},\ldots,\bm{T}_{\text{b}m})$.

\subsubsection{Permute the residuals under the modified model}\label{sec:QtY}

As discussed in the previous section, the residuals of the reduced model are asymptotically second moment exchangeable. 
 \cite{HuhJhun2001} and \cite{Renaud2010} discussed a further transformation, which will give second moment exchangeability.\\

Let $Q$ be a $n \times (n-d)$ matrix constructed from the eigenvectors of $(I_n-H_{\Lambda})$ such that $QQ^T=(I_n-H_{\Lambda})$ and $Q^TQ=I_{n-d}$. Multiplying both 
sides of Equation \eqref{model2} by $Q^T$ gives the modified model 
\begin{align}
Q^T\bm{Y} = Q^TX_{\text{g}}\bm{\beta}_{\text{g}}+Q^T\bm{\epsilon}. 
\label{model3}
\end{align}
Let $\tilde{\bm{Y}}=Q^T\bm{Y}, \tilde{X}_g=Q^TX_{\text{g}}$ and $\tilde{\bm{\epsilon}}=Q^T\bm{\epsilon}$. Then, 
\begin{align*}
\tilde{\bm{Y}} = \tilde{X}_g\bm{\beta}_{\text{g}} + \tilde{\bm{\epsilon}}. 
\end{align*}
Under $H_0: \bm{\beta}_{\text{g}}=\bm{0}$,
\begin{align*}
\text{E}(\tilde{\bm{Y}}) = \bm{0} \text{ and } \text{Cov}(\tilde{\bm{Y}}) =  \sigma^2I_{n-d}
\end{align*}
and the transformed data $\tilde{\bm{Y}}$ are second moment exchangeable.
If $\bm{Y}$ is assumed normally distributed, then $\tilde{\bm{Y}}$ is also exchangeable \citep{Solari2014}. 
The score test statistic for this model is given in Equation \eqref{Tk} and by permuting the
transformed responses, $\tilde{\bm{Y}}$, we can estimate the distribution of the maximal test statistic, and then estimate the local significance level, $\alpha_{\text{loc}}$.
Let $P$ be a $n \times n$ permutation matrix. The permuted score score test statistic for the genetic marker $j$ in the $b$'th permutation is
\begin{align}
\bm{T}_{\text{b}j} = \frac{X_{\text{g}j}^TP\tilde{\bm{Y}}}{\sqrt{X_{\text{g}j}^T\Lambda^{1/2}(I_n-H_{\Lambda})\Lambda^{1/2}X_{\text{g}j}}}.
\label{freedmanlane}
\end{align}
The vector of score test statistics for the $b$'th permutation is $\bm{T}_{\text{b}}=(\bm{T}_{\text{b}1},\ldots,\bm{T}_{\text{b}m})$.

\cite{Renaud2010}
proved that if the joint distribution of $\bm{\epsilon}$ is spherical, then the distribution of $\tilde{\bm{Y}}$ is also spherical and the elements of $\tilde{\bm{Y}}$ are exchangeable. If 
the distribution of $\tilde{\bm{Y}}$ is exchangeable and $P$ is a permutation matrix, we can obtain a permutation test controlling the FWER at level $\alpha$.
The method of \cite{Renaud2010} is based on permuting the residuals of the modified model, and is used in the ANOVA setting. 
\cite{HuhJhun2001} use the same type of approach in the regression case, but only in the case of univariate hypothesis testing. The paper of \cite{HuhJhun2001}
also discuss a multivariate test, but their multivariate approach is based on bootstrapping the residuals under the full model. 

\cite{Solari2014} used the rotation tests as described by \cite{Langsrud2005} in the context of multiple testing. Their method is also based on the modified model as in 
Equation \eqref{model3}, but instead of using permutation matrices $P$, they use rotation matrices, $R^{*}$. Permutation matrices, $P$, are a subset of all
possible rotation matrices $R^{*}$ satisfying $R^{*T}R^{*}=I_{n}$. \cite{Solari2014} also assume that the test statistics are multivariate normally distributed, and as proved by 
\cite{Renaud2010} this will give a permutation test controlling the FWER at level $\alpha$.

\subsubsection{Permute the residuals under the full model}\label{sec:fullmodel}

\cite{TerBraak1992} introduced permuting the residuals under the full model. 
We fit the full regression model, 
\begin{align*}
\bm{Y}=X_{\text{e}}\bm{\beta}_{\text{e}}+X_{\text{g}}\bm{\beta}_{\text{g}}+\bm{\epsilon},
\end{align*}
to obtain estimates $\hat{\bm{\beta}}_{\text{e}}$ of $\bm{\beta}_{\text{e}}$,  $\hat{\bm{\beta}}_{\text{g}}$ of $\bm{\beta}_{\text{g}}$ and the residuals $\hat{\bm{\epsilon}}^{*}$. We get the fitted values
\begin{align*}
\bm{Y}^{*}=X_{\text{e}}\hat{\bm{\beta}}_e+X_{\text{g}}\hat{\bm{\beta}}_g+\hat{\bm{\epsilon}}^{*}. 
\end{align*}
The method of \cite{TerBraak1992} is based on resampling without replacement from the residuals $\hat{\bm{\epsilon}}^{*}$. 
\cite{WestfallYoung1993} also discussed a regression-based resampling method, based on the residuals of the full model, 
but this method is based on resampling with replacement from $\hat{\bm{\epsilon}}^{*}$. 
We write the model as
\begin{align*}
(I_n-H_{\Lambda})\bm{Y}=(I_n-H_{\Lambda})X_{\text{e}}\bm{\beta}_{\text{e}} + (I_n-H_{\Lambda})X_{\text{g}}\bm{\beta}_{\text{g}} + (I_n-H_{\Lambda})\bm{\epsilon}, 
\end{align*}
and under the null hypothesis $H_0: \bm{\beta}_{\text{g}}=\bm{0}$, 
\begin{align*}
(I_n-H_{\Lambda})\bm{Y}=(I_n-H_{\Lambda})X_{\text{e}}\bm{\beta}_{\text{e}} + (I_n-H_{\Lambda})\bm{\epsilon}. 
\end{align*}
Under $H_0: \bm{\beta}_{\text{g}}=\bm{0}$, the expected value of the residuals are  $\text{E}[(I_n-H_{\Lambda})\bm{Y}]=(I_n-H_{\Lambda})X_{\text{e}}\bm{\beta}_{\text{e}}$, 
which in general is not exchangeable. The covariance matrix of the residuals is $\text{Cov}[(I_n-H_{\Lambda})\bm{Y}] = \sigma^2(I_n-H_{\Lambda})$. 
Let $P$ be a $n \times n$ permutation matrix. The permuted score score test statistic for the genetic marker $j$ in the $b$'th permutation is
\begin{align}
\bm{T}_{\text{b}j} = \frac{X_{\text{g}j}^TP\hat{\bm{\epsilon}}^{*}}{\sqrt{X_{\text{g}j}^T\Lambda^{1/2}(I_n-H_{\Lambda})\Lambda^{1/2}X_{\text{g}j}}}
\label{terbraak1992}
\end{align}
where $\hat{\bm{\epsilon}}^{*}$ are the residuals from the full model. The vector of score test statistics for the $b$'th permutation is $\bm{T}_{\text{b}}=(\bm{T}_{\text{b}1},\ldots,\bm{T}_{\text{b}m})$.

Permutation of the residuals under the full model can be seen as permutation under the alternative hypothesis. In Section \ref{sec:results}, we compare the different
permutation methods by the estimated FWER, that is, we consider methods where the permutation is done under the complete null hypothesis. The method of \cite{TerBraak1992}
is included in this section as an example of methods for permutation testing for the normal linear model, but not considered further in this paper. 

In this section, permutation methods for the normal linear model are presented. In Section \ref{sec:simulations}, we will use simulated data to compare the results based on using these permutation methods. 

\subsection{Permutation methods for generalized linear models}\label{sec:glm}
In this section we will present a new permutation method for generalized linear models.
For generalized linear models, the exchangeability assumption is in general not satisfied. 
Following \cite{Commenges2003} we aim to obtain second moment exchangeability by using a transformation, $\tilde{\bm{Y}}$, 
such that
\begin{align*}
\text{E}(\tilde{\bm{Y}})=a \text{ and } \text{Cov}(\tilde{\bm{Y}})=b\cdot I_n
\end{align*}
where $a$ and $b$ are constant values. If the values of $\bm{\mu}_{\text{e}}$ and $\text{Cov}(\bm{Y})=\Lambda$ were known, we could use the transformation
\begin{align*}
\tilde{\bm{Y}}=\Lambda^{-1/2}(\bm{Y}-\bm{\mu}_{\text{e}}), 
\end{align*}
which has
\begin{align*}
\text{E}(\tilde{\bm{Y}})=\Lambda^{-1/2}(\text{E}(\bm{Y})-\bm{\mu}_{\text{e}})=\bm{0}
\end{align*}
and
\begin{align*}
\text{Cov}(\tilde{\bm{Y}})=\Lambda^{-1/2}\text{Cov}(\bm{Y})\Lambda^{-1/2}=\Lambda^{-1/2}\Lambda \Lambda^{-1/2}=I_n. 
\end{align*}
This gives second moment exchangeability. 
For regression models, $\bm{\mu}_{\text{e}}$ and $\Lambda$ are in general unknown and need to be estimated.

\subsubsection{The $\Lambda$-method}
Based on the transformation presented above, we suggest a new method, which we call the $\Lambda$-method, for logistic regression models in combination with the maxT permutation method. 
The algorithm consists of the following steps,

\begin{enumerate}
	\item Let $\hat{\bm{\mu}}_{\text{e}}$ be the GLM estimate of $\bm{\mu}_{\text{e}}$.
	\item Define $\hat{\Lambda}=\text{diag}(\hat{\bm{\mu}}_{\text{e}i}(1+\hat{\bm{\mu}}_{\text{e}i}))$. 
	\item Construct $\bm{\tilde{Y}}=\hat{\Lambda}^{-1/2}(\bm{Y}-\hat{\bm{\mu}}_{\text{e}})$.
	\item Construct $\tilde{X}_{\text{g}j} = \frac{\hat{\Lambda}^{1/2}X_{\text{g}j}}{\sqrt{X_{\text{g}j}^T\hat{\Lambda}^{1/2}(I_n-H_{\hat{\Lambda}}) \hat{\Lambda}^{1/2} X_{\text{g}j}}}, j=1,\ldots, m$.
	\item Permute $\bm{\tilde{Y}}$ to yield $P\bm{\tilde{Y}}$. 
	\item For each permuted dataset $b=1,\ldots,B$ the permuted score test statistics for the $m$ genetic markers are $\bm{T}_{\text{b}j}=\tilde{X}_{\text{g}j}P\bm{\tilde{Y}}, j=1,\ldots, m$, where $P$ is a permutation matrix, Then, $\bm{T}_b=(T_{\text{b}1},\ldots,T_{\text{b}m})$
	\item Calculate $\max|\bm{T}|_b=\max(|T_{\text{b}1}|,\ldots,|T_{\text{b}m}|)$ for each permuted sample $b=1,\ldots,B$. Order the $B$ maximal test statistics as $\max|T_{(1)}| \leq \cdots \leq  \max|T_{(B)}|$. 
	\item We are interested in controlling the FWER at level $(1-q)$. The cutoff value for the maximal test statistic is given by element number $Bq$ in the sorted vector of the $B$ maximal test statistics, $Q=\max |T_{(Bq)}|$. Confidence interval for the cutoff-value is calculated as described in Appendix \ref{sec:ci}.
	\item The local significance level is given by $\alpha_{\text{loc}}=2(1-\Phi(Q))=2\Phi(-Q)$ assuming $\bm{T}_b$ is multivariate normally distributed. 
\end{enumerate}

The $\Lambda$-method is presented above for the logistic regression model, but only step 2 in the algorithm is dependent on the regression model. 
If we replace $\hat{\Lambda}$ with $\text{diag}(\hat{\sigma}_i^2)$, the $\Lambda$-method can be used also for other types of GLM, e.g. the Poisson GLM.

\subsection{Bootstrap methods}\label{sec:bootstrap}

Bootstrap methods do not depend on the assumption of exchangeability. In Section \ref{sec:simulations} we will use simulated data to compare our 
permutation method with parametric bootstrap for generalized linear models with normal or binomial distributed response variable. The parametric bootstrap
method sample with replacement from the estimated parametric distribution of the data. For the binomial model, the expected 
value, $\bm{\mu}_e$, need to be estimated, and for the normal linear model, $\text{Var}(Y_i)=\sigma^2$ need to be estimated. 
Therefore, the bootstrap $p$-values will only be asymptotically valid.

\section{Results}\label{sec:results}
In this section we use simulated data to compare and evaluate the different methods presented in Section \ref{sec:resamplingmethods}.  
We compare the methods using the estimated local significance level and the estimated FWER. We also include the method of permuting the raw data, $\bm{Y}$.

\subsection{Simulations}\label{sec:simulations}

We simulate genetic markers with alleles $A$ and $a$, where $A$ is assumed to be the high risk allele. The $A$ allele is coded as $1$ and the $a$ allele is coded as $0$. The minor allele frequencies (MAF) for the genetic markers are simulated from a uniform distribution on the interval $[0.05,0.5]$. We have $P(A)=\text{MAF}$ and $P(a)=1-\text{MAF}$. The combination of the two alleles at a given position on the DNA gives the genotype, $aa$, $Aa$ or $AA$, coded as $0$, $1$ or $2$, respectively. We simulate data for $m$ correlated genetic markers based on a latent multivariate normally distributed variable with a given correlation matrix, $\Sigma$, for example a matrix of compound symmetry correlation structure. The singular value decomposition of $\Sigma$ is $\Sigma=UDV^T$ and we denote $\Sigma_1 = UD^{1/2}$. We also simulated one environmental covariate following a standard normal distribution, $X_e \sim N(0,1)$ with effect size $\beta_e$. 

The genetic markers were simulated using the following algorithm (with inspiration from \cite{simSNP}). Each individuals two copies of the DNA are simulated independently of each other. 
\begin{enumerate}
\item Start by simulating a multivariate normally distributed variable, $X_0 \sim \text{N}_m(0,I)$ and calculate $X_1 = \Sigma_1X_0$. $X_1$ is multivariate normally distributed $X_1 \sim \text{N}_m(0,\Sigma)$.  
\item Then, we calculate $W=\Phi^{-1}(\text{MAF})$ and dichotomize the variable $X_1$ with $1$ if $X_1 < W$ and $0$ if $X_1 > W$, giving a vector $X_2$ of $0$'s and $1$'s, representing the alleles on one copy of the DNA. 
\item We simulate alleles for the second copy of the DNA similarly and independently of $X_2$, giving a vector denoted by $X_3$. 
\item The genotype for each of the $m$ genetic markers are found by $X_4=X_2+X_3$. 
\item Repeat 1-4 to give the genotype data for the $n$ individuals. 
\end{enumerate} 

We simulated $K$ independent datasets and each dataset was resampled $B$ times. The number of genetic markers is $m=100$, the number of simulated datasets is K=$1000$ or K=$5000$ and the number of permutations or bootstraps of each simulated dataset is $B=1000$ in all simulations considered. For permutation methods based on the maximal test statistic and random permutations, \cite{Goeman2014} write that $1000$ permutations is usually sufficient at $\alpha=0.05$, independent of the number $m$ of genetic markers. The data were simulated based on a latent multivariate normal variable with a compound symmetry correlation structure with correlation coefficient $\rho=0.7$. With correlation $\rho=0.7$ as input to the simulation code, the mean correlation coefficient in the correlation matrix of $1000$ simulated data sets varies between $0.3667$ and $0.4768$. How to simulate SNPs with a given correlation structure is not considered further in this paper. 

The different resampling methods were compared using the estimated FWER. We simulated independent datasets and applied different methods presented in Section \ref{sec:resamplingmethods} to estimate the local significance level, $\alpha_{\text{loc}}$. We estimated the FWER in each simulated dataset as
\begin{align*}
\hat{\alpha}_k=\frac{\#(\max |\bm{T}_b| \geq \max |\bm{t}_{\text{org}}|)+1}{B+1}, k=1,\ldots,K
\end{align*}
where $\bm{T}_b$ is the test statistics from the $b$'th resampled dataset, $\bm{t}_{\text{org}}$ is the observed test statistics from the simulated dataset and $B$ is the number of permutations/bootstraps of each dataset. This gives $K$ estimated FWER values, $\hat{\alpha}_1,\ldots,\hat{\alpha}_K$. We estimated the FWER for each of the resampling methods by
the proportion of simulated datasets with at least one false positive result,  
\begin{align*}
\tilde{\alpha}=\frac{\#(\hat{\alpha}_k \leq \alpha)}{K}
\end{align*}
where $\alpha=0.05$ and calculated a $95\%$ confidence interval for the estimated FWER, $\tilde{\alpha}$, as
\begin{align*}
\left[\tilde{\alpha} \pm 1.96\sqrt{\frac{\tilde{\alpha}(1-\tilde{\alpha})}{K}}\right].
\end{align*}
We also compared the local significance level for the different methods with the numerical integration method by \cite{genz1992,genz1993} which is implemented for $m \leq 1000$ in the R package mvtnorm \citep{mvtnorm}. This method can be used to solve the high dimensional integral in Equation \eqref{fwer} with a given value of precision for arbitrary correlation matrices. There exists different types of confidence intervals that can be calculated for the maximal test statistics, we calculate the confidence intervals for the maximal test statistic as described in Appendix \ref{sec:ci}.

\subsubsection{Normal linear model}

In this section we present results for some permutation and bootstrap methods using a normal linear regression model. The number of simulated datasets is $K=5000$ for the normal linear regression model. The $Y$ method is based on permuting the raw data, $\bm{Y}$.

Table \ref{simfwer4000} shows the estimated FWER using different resampling methods for simulated data with $m=100$ genetic markers, $n=400$ individuals and different values of $\beta_{\text{e}}$. 
For $\beta_{\text{e}}=0.0$ we see that the method based on permuting the raw data, $\bm{Y}$, gives estimated FWER level with confidence interval including $0.05$, as expected since the exchangeability
assumption is satisfied. We also see that the permutation method based on permuting the raw data, $\bm{Y}$, is conservative for $\beta_{\text{e}}>0.0$, while the other methods control the FWER at level $\alpha=0.05$.

\begin{table}[h!]
\begin{center}
\vspace{0.5cm}
\begin{tabular}{lccc}
Method				& $\beta_{\text{e}}$		& $\hat{\alpha}$ & $95\%$ C. I.  \\
\hline 
\cite{FreedmanLane1983},  The $\Lambda$-method  & $0.0$			& $0.0522$ 		& ($0.0460$, $0.0584$) \\
\cite{Renaud2010}			        & $0.0$		& $0.0502$ 		& ($0.0441$, $0.0563$) \\
Y       						& $0.0$		& $0.0516$ 		& ($0.0455$, $0.0577$) \\
Bootstrap    					& $0.0$		& $0.0480$ 		& ($0.0421$, $0.0539$) \\
\hline
\cite{FreedmanLane1983}, The $\Lambda$-method   & $0.5$			& $0.0522$ 		& ($0.0460$, $0.0584$) \\
\cite{Renaud2010}      			& $0.5$		& $0.0502$ 		& ($0.0441$, $0.0563$) \\
Y        						& $0.5$		& $0.0158$ 		& ($0.0123$, $0.0193$) \\
Bootstrap    					& $0.5$		& $0.0474$ 		& ($0.0415$, $0.0533$) \\
\hline
\cite{FreedmanLane1983}, The $\Lambda$-method   & $1.0$			& $0.0522$ 				& ($0.0460$, $0.0584$) \\
\cite{Renaud2010}      			& $1.0$		& $0.0502$ 				& ($0.0441$, $0.0563$) \\
Y        						& $1.0$		& $0.0002$ 				& ($0.0000$, $0.0006$) \\ 
Bootstrap    					& $1.0$		& $0.0484$ 				& ($0.0425$, $0.0543$) \\
\end{tabular}
\caption{Estimated FWER using simulated normally distributed data ($K=5000$ simulated datasets, $B=1000$ permutations/bootstraps of each dataset). } 
\label{simfwer4000}
\end{center}
\end{table}

\subsubsection{Binomial GLM}

In this section we present results for the estimated FWER using different resampling methods for the binomial GLM (logistic regression). 
When $\beta_{\text{e}}=0.0$, the exchangeability assumption is satisfied, that is, the method based on permuting the raw data, $\bm{Y}$, will control the FWER at level $\alpha=0.05$. 

Table \ref{simfwerbin} shows the estimated FWER for simulated data with different values of $\beta_{\text{e}}$.
From Table  \ref{simfwerbin} we see that the method based on permuting the raw data, $\bm{Y}$, is conservative when $\beta_{\text{e}}>0$ for the parameters in the simulation study. For $\beta_{\text{e}}=0.0$, the method permuting $\bm{Y}$ controls the FWER at level $\alpha=0.05$. 
From Table \ref{simfwerbin}, we also see that the $\Lambda$-method is conservative when $\beta_{\text{e}}>0$. The bootstrap method controls the FWER at level $\alpha=0.05$ in all examples considered. 
We also compared the
different methods using simulated data with $n=2000$ individuals. Table \ref{simfwerbin} shows that the results using the different methods are similar for $\beta_{\text{e}}=1.5$ using $n=400$ or $n=2000$ individuals. 

\begin{table}[h!]
\begin{center}
\vspace{0.5cm}
\begin{tabular}{lcccc}
Method				& $\beta_{\text{e}}$		& $n$	& $\hat{\alpha}$ & $95\%$ C. I.  \\
\hline  
The $\Lambda$-method	& $0.0$	 & $400$	& $0.041$		& ($0.0287$, $0.0533$) \\ 
Y					& $0.0$	 & $400$	& $0.042$		& ($0.0296$, $0.0544$) \\ 
Bootstrap 				& $0.0$	 & $400$	& $0.045$		& ($0.0322$, $0.0578$) \\ 
\hline
The $\Lambda$-method	& $1.5$ 	 & $400$ & $0.032$		& ($0.0211$, $0.0429$) \\ 
Y					& $1.5$	 & $400$ & $0.006$ 		& ($0.0012$, $0.0108$) \\ 
Bootstrap				& $1.5$	 & $400$ & $0.047$		& ($0.0339$, $0.0601$) \\ 
\hline
The $\Lambda$-method	& $1.5$	 	& $2000$	& $0.034$		& ($0.0228$, $0.0452$) \\
Y					& $1.5$		& $2000$	& $0.008$		& ($0.0025$, $0.0135$) \\ 
Bootstrap				& $1.5$		& $2000$	& $0.056$		& ($0.0417$, $0.0703$) \\
\end{tabular}
\caption{Estimated FWER using simulated binomial distributed data ($K=1000$ simulated datasets, $B=1000$ permutations/bootstraps of each dataset). } 
\label{simfwerbin}
\end{center}
\end{table}

\subsection{The local significance level}
Table \ref{simgaussian} shows the estimated local significance level using simulated data with $m=100$ genetic markers from a normal linear model using the different resampling methods presented in this paper. The effect size of the environmental covariate was $\beta_{\text{e}}=1.5$, and 
the environmental covariate was standard normally distributed, $X_{\text{e}} \sim N(0,1)$. The sample size was $n=400$. The simulated data were permuted $B=5000$ times since the aim was to estimate the local significance level, $\alpha_{\text{loc}}$ using Equation \eqref{c}. In addition, since $m<1000$ we also calculated the local significance level using the numerical integration method by \cite{genz1992,genz1993}, giving $\alpha_{\text{loc}}=0.0007998471$.  The exchangeability assumption was not satisfied since $\beta_{\text{e}}>0$, and from Table \ref{simgaussian} we see that the method permuting the raw data, $\bm{Y}$, is conservative, i.e. the value of $\alpha_{\text{loc}}$ is lower than using the method by \cite{genz1992,genz1993}. 

\begin{table}[h!]
\begin{center}
\vspace{0.5cm}
\begin{tabular}{lr}
Method 				& $\alpha_{\text{loc}}$	\\ 	
\hline
Y					& $6.6640 \cdot 10^{-10}$			\\
Freedman and Lane		& $0.0007985001$				\\
Bootstrap				& $0.0008046713$				\\
The $\Lambda$-method	& $0.0008131796$				\\
\end{tabular}
\caption{Estimated $\alpha_{\text{loc}}$ using simulated data from a normal linear model with $m=100$ genetic markers.}
\label{simgaussian}
\end{center}
\end{table}

Table \ref{simbinomial} shows the estimated local significance level using simulated data with $m=100$ independent genetic markers from a binomial GLM. The effect size of the environmental covariate was $\beta_{\text{e}}=1.5$ and $X_{\text{e}} \sim N(0,1)$.
The sample size was $n=400$ and the data was permuted $B=5000$ times to estimate $\alpha_{\text{loc}}$ using Equation \eqref{c}. In addition, since $m<1000$ we also calculated the local significance level using the numerical integration method by \cite{genz1992,genz1993}, giving $\alpha_{\text{loc}}=0.000818057$.  The exchangeability assumption was not satisfied since $\beta_{\text{e}}>0$, and from Table \ref{simgaussian} we see that the method permuting the raw data, $\bm{Y}$, is conservative, i.e. the value of $\alpha_{\text{loc}}$ is lower than using the method by \cite{genz1992,genz1993}.

\begin{table}[h!]
\begin{center}
\vspace{0.5cm}
\begin{tabular}{lr}
Method 				& $\alpha_{\text{loc}}$	\\ 	
\hline
Y					& $4.6616 \cdot 10^{-5}$				\\
The $\Lambda$-method	& $0.000813180$				\\
Bootstrap				& $0.000829454$				\\
\end{tabular}
\caption{Estimated $\alpha_{\text{loc}}$ using simulated data from a binomial GLM with $m=100$ genetic markers.}
\label{simbinomial}
\end{center}
\end{table}

\section{Discussion}\label{sec:discussion}

In this paper we have presented and discussed resampling methods for generalized linear models. Methods for permutation testing in the normal linear model are reviewed and compared using simulated data, and used in the context of multiple testing. We have also discussed permutation testing for GLMs, and the concept of exchangeability for regression models. 

For the normal linear model, $\text{E}(\bm{Y})=X_e\bm{\beta_{\text{e}}}$ and when $\beta_{\text{e}}\neq 0$, the observations, $Y_i$, will in general have different expected values, $\text{E}(Y_i)=X_{\text{e}i}\beta_{e}$, and the exchangeability assumption is in general not satisfied. There exists approximate methods for permutation testing in the normal linear model as presented in Section \ref{sec:lm}, including the methods of \cite{FreedmanLane1983} and \cite{Renaud2010}. In this paper, these methods are used in the multiple testing setting. 

For a logistic regresion model, $\text{E}(Y_i)=\frac{\exp(X_{\text{e}i}\beta_{\text{e}})}{1+\exp(X_{\text{e}i}\beta_{\text{e}})}$ and when $\beta_{\text{e}} \neq 0$, the exchangeability assumption is in general not satisfied and to our knowledge, there is no standard solution on how to do permutation testing, except when the model includes only discrete covariates. For a model including only discrete covariates, the exchangeability assumption can be satisfied by permuting the data, $\bm{Y}$, within subgroups of the environmental covariate \citep{Solari2014}.

Another strategy for permutation testing in GWA studies have been employed by e.g. 
\cite{ConneelyBoehnke2007}. They permuted the individual genotype vectors while the environmental covariate and phenotypes were not permuted. This method will change the correlation between the environmental covariate and the genotypes and can therefore not be used when the environmental covariate is for example population structure. Population structure can be adjusted for by including principal components of the genotype correlation matrix as environmental covariates \citep{Price2006}, but then the environmental covariate and the genotypes are correlated. 

In Section \ref{sec:glm} we presented an alternative method for permutation testing in GLMs which can be used both when the response variable is binomial or normally distributed. The method is described in a multiple testing setting and named the $\Lambda$-method. For the normal linear model, the $\Lambda$-method is equivalent to the method of \cite{FreedmanLane1983}. The $\Lambda$-method can be used for both discrete and continuous environmental covariates. 

We used simulated data to compare the $\Lambda$-method to other resampling methods. The data were simulated under the complete null hypothesis of no association between the genetic markers and the phenotype. The resampling methods were compared in a multiple testing setting by the estimated FWER. We varied the sample size and the effect size of the environmental covariate. For the normal linear model we compared the $\Lambda$-method with the method of \cite{FreedmanLane1983}, \cite{Renaud2010}, the method permuting the raw data, $\bm{Y}$, and the parametric bootstrap method. For the binomial GLM we compared the $\Lambda$-method by the method permuting the raw data, $\bm{Y}$, and the parametric bootstrap method.  

For the normal linear model, the results of the simulations in Section \ref{sec:simulations} show that for sample size $n=400$ and our choice of simulation parameters, the $\Lambda$-method and the method of \cite{FreedmanLane1983} methods control the FWER at level $\alpha=0.05$. We also see that the method of \cite{Renaud2010} controls the FWER at level $\alpha=0.05$ in our examples. 

The method based on permuting the raw data, $\bm{Y}$, ignores the relationship between the response variable and the environmental covariate. The results of the simulations show that when the effect size of the environmental covariate increases (and for our choice of simulation parameters), this method becomes very conservative, both for the normal linear model and the logistic regression model. From the simulations in this paper, we also see that the $\Lambda$-method is conservative for $\beta_{\text{e}}>0$ in our examples for the binomial GLM. For the binomial model, we considered sample sizes $n=400$ and $n=2000$. 

The results of the simulation study show that the parametric bootstrap method controls the FWER at level $\alpha=0.05$ in all examples considered. The parametric bootstrap method makes an assumption about the distribution of the response based on estimated parameters from the original data, while the $\Lambda$-method only makes an assumption about the variance of the response variable. 

In this paper, we have presented different resampling methods for multiple testing correction and control of the FWER. 
We have also discussed the assumption of exchangeability, both for normal linear models and generalized linear models
and presented a new method for permutation in generalized linear models. The new method can be used when the 
environmental covariates are discrete or continuous and for different types of GLMs. 

However, further work is needed, in particular to assess the exchangeability assumption for the $\Lambda$-method and to assess and compare the power of the different methods. In \cite{Halle2016} we defined the most powerful method for multiple testing correction as the method
which yields the largest value of the local significance level, $\alpha_{\text{loc}}$, and controls the FWER at level $\alpha$.


\section*{Software}
The statistical analysis were performed using the statistical software R \citep{R}. R code used for the simulations in this paper are available at \\http://www.math.ntnu.no/$\sim$karikriz/resampling. 

\section*{Acknowledgements}
The authors would like to thank Dr. Brenton Clarke (Murdoch University, Perth, Western Australia) for valuable comments.  \\
The PhD position of the first author is founded by the Liaison Committee between the Central Norway Regional Health Authority (RHA) and the 
Norwegian University of Science and Technology (NTNU). 

\emph{Conflict of Interest:} None declared.

\bibliographystyle{chicago}
\bibliography{resampling}

\appendix


\section{The hat matrix for regression models}\label{sec:hatmatrix}

\subsection{The hat matrix for the linear regression model}
The hat matrix for the linear regression model is
 $H=X_{\text{e}}(X_{\text{e}}^TX_{\text{e}})^{-1}X_{\text{e}}^T$, where $X_{\text{e}}$ is the matrix of environmental  covariates, with the intercept in the first column. In Section \ref{sec:lm} 
methods for permutation testing are described, and some of the methods are based on the projection matrix $(I-H)$.

\subsubsection{Only intercept}
First consider the case with no environmental covariates. Then $X_{\text{e}}=\bm{1}_n$ and
element $(i,j)$ of the hat matrix is
\begin{align*}
H_{ij} = \frac{1}{n}. 
\end{align*}
We have $H_{ij} \rightarrow 0$ as $n \rightarrow \infty$. The diagonal elements of the projection matrix $M=(I-H)$ will then be
$M_{ii}=1-\frac{1}{n}, i = 1,\ldots, n$ and the off-diagonal elements willl be $M_{ij}=-\frac{1}{n}, (i,j)=1,\ldots, n$, so the matrix
M is of compound symmetry structure.

\subsubsection{One covariate}
Now consider one environmental covariate, $\bm{z}$, in addition to the intercept. We assume that the covariate is standardized such that $E(\bm{z})=\bm{0}$ and $\text{Var}(\bm{z})=I$. 
The matrix $X_{\text{e}}=[\bm{1}_n \hspace{0.1cm} \bm{z}]$ has two columns.  

We assume that we have one covariate, $\bm{z}$, in addition to the intercept, so $X_{\text{e}}=[\bm{1}_n \hspace{0.1cm} \bm{z}]$. We assume that the covariate is standardized such that $E(\bm{z})=\bm{0}$ and $\text{Var}(\bm{z})=I$. The hat matrix is given by 
\begin{align*}
H &= X_{\text{e}}(X_{\text{e}}^TX_{\text{e}})^{-1}X_{\text{e}}^T \\
	&= \begin{bmatrix}
       1 		 & z_1          \\[0.3em]	
       \vdots 	        & \vdots \\[0.3em]
       1            & z_n
     \end{bmatrix}
	\big(\begin{bmatrix}
	1 & \ldots & 1 \\[0.3em]
	z_1 & \ldots & z_n 
	\end{bmatrix}
	\begin{bmatrix}
       1 		 & z_1          \\[0.3em]	
       \vdots 	        & \vdots \\[0.3em]
       1            & z_n
     \end{bmatrix}
	\big)^{-1}
\begin{bmatrix}
	1 & \ldots & 1 \\[0.3em]
	z_1 & \ldots & z_n 
	\end{bmatrix} \\
&= \begin{bmatrix}
       1 		 & z_1          \\[0.3em]	
       \vdots 	        & \vdots \\[0.3em]
       1            & z_n
     \end{bmatrix}
	\big(\begin{bmatrix}
	n &  \sum_{i=1}^n z_i \\[0.3em]
	\sum_{i=1}^n z_i  & \sum_{i=1}^n z_i^2
	\end{bmatrix}
	\big)^{-1}
\begin{bmatrix}
	1 & \ldots & 1 \\[0.3em]
	z_1 & \ldots & z_n 
	\end{bmatrix}  \\
&= 	\frac{1}{n\sum_{i=1}^n z_i^2-(\sum_{i=1}^n z_i)^2} 
 \begin{bmatrix}
       1 		 & z_1          \\[0.3em]	
       \vdots 	        & \vdots \\[0.3em]
       1            & z_n
     \end{bmatrix}
\begin{bmatrix}
\sum_{i=1}^n z_i^2 & -\sum_{i=1}^n z_i\\[0.3em] 
-\sum_{i=1}^n z_i & n  
\end{bmatrix}
\begin{bmatrix}
	1 & \ldots & 1 \\[0.3em]
	z_1 & \ldots & z_n 
	\end{bmatrix}  \\
&= \frac{1}{n\sum_{i=1}^n z_i^2-(\sum_{i=1}^n z_i)^2} 
\begin{bmatrix}
\sum_{i=1}^n z_i^2-z_1\sum_{i=1}^n z_i & -\sum_{i=1}^n z_i+z_1n \\[0.3em]
\vdots & \vdots \\[0.3em]
\sum_{i=1}^n z_i^2-z_n\sum_{i=1}^n z_i & -\sum_{i=1}^n z_i+z_nn
\end{bmatrix}
\begin{bmatrix}
	1 & \ldots & 1 \\[0.3em]
	z_1 & \ldots & z_n 
	\end{bmatrix}  \\
\end{align*}
The covariate $\bm{z}$ are centered and standardized. This gives
\begin{align*}
H &= \frac{1}{n\sum_{i=1}^n z_i^2-(\sum_{i=1}^n z_i)^2} 
\begin{bmatrix}
n & z_1n \\[0,3em]
\vdots & \vdots \\[0.3em]
n & z_nn 
\end{bmatrix}
\begin{bmatrix}
	1 & \ldots & 1 \\[0.3em]
	z_1 & \ldots & z_n 
	\end{bmatrix}  \\
&= \frac{1}{n\sum_{i=1}^n z_i^2-(\sum_{i=1}^n z_i)^2} 
\begin{bmatrix}
n+z_1^2n & n+z_1z_2n & \cdots & n+z_1z_nn \\[0.3em]
\vdots & \ddots & \cdots & \vdots \\[0.3em]
n+z_nz_1n & n+z_nz_2n & \cdots & n+z_n^2n 
\end{bmatrix} \\
&= \frac{1}{n(\sum_{i=1}^n z_i^2-n\bar{z}^2)}
\begin{bmatrix}
n+z_1^2n & n+z_1z_2n & \cdots & n+z_1z_nn \\[0.3em]
\vdots & \ddots & \cdots & \vdots \\[0.3em]
n+z_nz_1n & n+z_nz_2n & \cdots & n+z_n^2n 
\end{bmatrix} \\
&= \frac{1}{n^2}
\begin{bmatrix}
n+z_1^2n & n+z_1z_2n & \cdots & n+z_1z_nn \\[0.3em]
\vdots & \ddots & \cdots & \vdots \\[0.3em]
n+z_nz_1n & n+z_nz_2n & \cdots & n+z_n^2n 
\end{bmatrix} \\
&= \frac{1}{n}
\begin{bmatrix}
1+z_1^2 & 1+z_1z_2 & \cdots & 1+z_1z_n \\[0.3em]
\vdots & \ddots & \cdots & \vdots \\[0.3em]
1+z_nz_1 & 1+z_nz_2 & \cdots & 1+z_n^2 
\end{bmatrix}
\end{align*}
The elements of the hat matrix are then given by
\begin{align*}
H_{ij}	&= \frac{1}{n}(1+z_iz_j) 
\end{align*}
We assume that the data contain no leverage points \citep[p.~207]{Brenton} and that the sample size
is large $z_iz_i << n$ for all values $i=1,\ldots,n$. Then, $H_{ii} \overset{n \rightarrow \infty}{\rightarrow} 0$. 
The off-diagonal elements of the hat matrix are bounded by $H_{ij} < H_{ii}(1-H_{ii})$. Then, 
$H_{ij} \overset{n \rightarrow \infty}{\rightarrow} 0$ for all values $i,j=1,\ldots,n$. This gives
$(I-H) \overset{n \rightarrow \infty}{\rightarrow} I$.

\section{Estimated confidence interval for the maximal test statistic}\label{sec:ci}
Let $Z$ be a random variable with cumulative distribution function $F_Z$. For a given $q$, let
\begin{align*}
P(Z < z_q) = q,
\end{align*}
where the quantile $z_q$ is the parameter of interest. 
We have observed a random sample of size $B$ from $F_Z$, $z_1,\cdots,z_B$. An estimator for the quantile $z_q$ is $Q$, where
$Q$ is the value of the $(q \cdot B)$'th order statistic in the sample. 

We are interested in a $(1-\alpha)\cdot 100\%$ confidence interval for $z_q$. This confidence interval can be found as follows: 

Let $W$ be the number of observations in our sample that is smaller than $z_q$. Then,
\begin{align*}
W \sim \text{Bin}(B,q). 
\end{align*}
Let $z_{(1)},\cdots,z_{(B)}$ be the ordered observations from the random sample.  Then, 
the event $(z_{(i)}<z_q)$ is identical to the event $(w \geq i)$ and $(z_{(i)}>z_q)$ is identical to the event $(w < i)$. 

Thus, for elements $(r,s) \in \{1,\cdots,B \}$, we have
\begin{align*}
P(z_{(r)} < Z_q < z_{(s)}) = P(r \leq W \leq s). 
\end{align*}
We choose $r,s$ such that 
\begin{align*}
P(r \leq W \leq s) = 1-\alpha. 
\end{align*}
This can be done numerically by finding $\delta$ where $r=Bq-\delta$ and $s=Bq+\delta$. 
As a result we have
\begin{align*}
Q=Z_{[Bq]}
\end{align*}
as the estimator for $z_q$ and confidence limits $z_{(r)}$ and $z_{(s)}$ where $[Bq]$ is the element number $Bq$ in vector $Z$.   

\end{document}